\documentstyle[amssymb,12pt]{article}

\input{tcilatex}

\begin{document}

\title{An improved phase error tolerance in quantum search algorithm}
\author{Jin-Yuan Hsieh$^{1}$, Che-Ming Li$^{2}$, and Der-San Chuu$^{2}$ \\
$^{1}$.Department of Mechanical Engineering, Ming Hsin University \\
of Science and Technology, Hsinchu 30401, Taiwan.\\
$^{2}$Institute and Department of Electrophysics, National Chiao \\
Tung University, Hsinchu 30050, Taiwan.\\
}
\maketitle

\begin{abstract}
As the matching condition in Grover search algorithm is transgressed due to
inevitable errors in phase inversions, it gives a reduction in maximum
probability of success. With a given degree of maximum success, we have
derive the generalized and imroved criterion for tolerated error and
corresponding size of quantum database under the inevitable gate
imperfections. The vanished inaccurancy to this condition has also been
shown. Besides, a concise formula for evaluating minimum number of
iterations is also presented in this work.

PACS: 03.67.Lx
\end{abstract}

Grover's quantum search algorithm\cite{Grover} provides a quadratic speedup
over classical counterpart, and it has been proved to be optimal for
searching a marked element with minimum oracle calls\cite{Zalka}. It is
achieved by applying Grover kernel on an uniform superposition state, which
is obtained by applying Walsh-Hadamard transformation on a initial state, in
a specific operating steps such that the probability amplitude of marked
state is amplified to a desired one. Grover's kernel is composed of phase
rotations and Walsh-Hadamard transformations. The phase rotations include
two kinds of operations : $\pi $-inversion of the marked state and $\pi $%
-inversion of the initial state. It has shown that the phases, $\pi $, can
be replaced by two angles, $\phi $ and $\theta $, under the phase matching
criterion, which is the necessary condition for quantum searching with
certainty. In other words, the relation between $\phi $ and $\theta $ will
affect the degree of success of quantum search algorithm. There have been
several studies concern with the effect of imperfect phase rotations. In
their paper\cite{Long 1}, Long et al. have found that the tolerated angle
difference between two phase rotations, $\delta $, due to systematic errors
in phase inversions, with a given expected degree of success $P_{\max }$, is
about $2/\sqrt{NP_{\max }}$, where $N$ is the size of the database. H{\small %
\O }yer\cite{Hoyer} has shown that after some number of iterations of Grover
kernel, depending on $N$ and unperturbed $\theta $, it will give a solution
with error probability $O(1/N)$ under a tolerated phase difference $\delta
\backsim O(1/\sqrt{N})$. The same result is also derived by Biham et al.\cite%
{Biham}. On the other hand, a near conclusion, $\delta \backsim O(1/N^{2/3})$%
, is presented by Pablo-Norman and Ruiz-Altaba\cite{Pablo-Norman}.

The result of Long et al\cite{Long 1} is based on the approximate Grover
kernel and an assumption: large $N$ and small $\delta $ et al. However, we
found that the main inaccuracy comes from the approximate Grover kernel.
Since all parameters in Grover kernel connect with each other exquisitely,
any reduction to the structure of Grover's kernel would destroy this
penetrative relation, so accumulative errors emerge from the iterations to a
quantum search. Although this assumption lead their study to a proper
result, it cannot be applied to general cases, e.g. any set of two angles in
phase rotations satisfies phase matching condition\cite{Long 2}\cite{hsieh
and li}. In what follows, we will get rid of the approximation to Grover
kernel, then derive an improved criterion for tolerated error in phase
rotation and the required number of qubits for preparing a database.
Besides, a concise formula for evaluating minimum number of iterations to
achieve a maximum probability will also be acquired. By this formula then
evaluating the actual maximum probability, one can realize the derived
criterion for tolerated error is near exactly.

The Grover kernel is composed of two unitary operators $G_{\tau }$ and $%
G_{\eta }$, given by%
\begin{eqnarray}
G_{\tau } &=&I+(e^{i\phi }-1)\left| \tau \right\rangle \left\langle \tau
\right| \text{,} \\
G_{\eta } &=&I+(e^{i\theta }-1)W\left| \eta \right\rangle \left\langle \eta
\right| W^{-1}\text{ ,}  \nonumber
\end{eqnarray}%
where $W$ is Walsh-Hadamard transformation, $\left| \tau \right\rangle $ is
the marked state, $\left| \eta \right\rangle $ is the initial state, and $%
\phi $ and $\theta $ are two phase angles. It can also be expressed in a
matrix form as long as an orthonormal set of basis vectors is chosen. The
orthonormal set is 
\begin{equation}
\left| I\right\rangle =\left| \tau \right\rangle \text{ and }\left| \tau
_{\perp }\right\rangle =(W\left| \eta \right\rangle -W_{\tau \eta }\left|
\tau \right\rangle )/l\text{ ,}
\end{equation}%
where $W_{\tau \eta }=\left\langle \tau \right| W\left| \eta \right\rangle $
and $l=(1-\left| W_{\tau \eta }\right| ^{2})^{1/2}$. Letting $W_{\tau \eta
}=\sin (\beta )$, we can write, from (2),

\begin{equation}
\left| s\right\rangle =W\left| \eta \right\rangle =\sin (\beta )\left| \tau
\right\rangle +\cos (\beta )\left| \tau _{\perp }\right\rangle \text{ ,}
\end{equation}%
and Grover kernel can now be written%
\begin{eqnarray}
G &=&-\text{ }G_{\eta }G_{\tau }  \nonumber \\
&=&-\left[ 
\begin{array}{cc}
e^{i\phi }(1+(e^{i\theta }-1)\sin ^{2}(\beta )) & (e^{i\theta }-1)\sin
(\beta )\cos (\beta ) \\ 
e^{i\phi }(e^{i\theta }-1)\sin (\beta )\cos (\beta ) & 1+(e^{i\theta
}-1)\cos ^{2}(\beta )%
\end{array}%
\right] \text{.}
\end{eqnarray}

After $m$ number of iterations, the operator $G^{m}$ can be expressed as

\begin{equation}
G^{m}=(-1)^{m}e^{im(\frac{\phi +\theta }{2})}\left[ 
\begin{array}{cc}
e^{imw}\cos ^{2}(x)+e^{-imw}\sin ^{2}(x) & e^{-i\frac{\phi }{2}}i\sin
(mw)\sin (2x) \\ 
e^{i\frac{\phi }{2}}i\sin (mw)\sin (2x) & e^{imw}\sin ^{2}(x)+e^{-imw}\cos
^{2}(x)%
\end{array}%
\right] \text{.}
\end{equation}%
where the angle $w$ is defined by

\begin{equation}
\cos (w)=\cos (\frac{\phi -\theta }{2})-2\sin (\frac{\phi }{2})\sin (\frac{%
\theta }{2})\sin ^{2}(\beta )\text{ ,}
\end{equation}%
or

\begin{equation}
\sin (w)=\sqrt{(\sin (\frac{\theta }{2})\sin (2\beta ))^{2}+(\sin (\frac{%
\phi -\theta }{2})+2\sin (\frac{\theta }{2})\cos (\frac{\phi }{2})\sin
(\beta ))^{2}}\text{,}
\end{equation}%
the angle $x$ is defined by

\begin{equation}
\sin (x)=\sin (\frac{\theta }{2})\sin (2\beta )/\sqrt{l_{m}},
\end{equation}%
where

\begin{eqnarray*}
l_{m} &=&(\sin (w)+\sin (\frac{\phi -\theta }{2})+2\cos (\frac{\phi }{2}%
)\sin (\frac{\theta }{2})\sin ^{2}(\beta ))^{2}+(\sin (\frac{\theta }{2}%
)\sin (2\beta ))^{2} \\
&=&2\sin (w)(\sin (w)+\sin (\frac{\phi -\theta }{2})+2\cos (\frac{\phi }{2}%
)\sin (\frac{\theta }{2})\sin ^{2}(\beta ))\text{.}
\end{eqnarray*}%
More details can be found in the study\cite{hsieh and li}. Then the
probability of finding a marked state is

\begin{eqnarray}
P &=&1-\left| \left\langle \tau _{\perp }\right| G^{m}\left| s\right\rangle
\right| ^{2} \\
&=&1-(\cos (mw)\cos (\beta )-\sin (mw)\sin (\frac{\phi }{2})\sin (2x)\sin
(\beta ))^{2}  \nonumber \\
&&-\sin ^{2}(mw)(\cos (\frac{\phi }{2})\sin (2x)\sin (\beta )-\cos (2x)\cos
(\beta ))^{2}\text{.}  \nonumber
\end{eqnarray}%
Moreover, by the equation $\partial P/\partial (\cos (mw))=0$, the minimum
number of iterations for obtaining the maximum probability, $P_{\max }(\cos
(m_{\min }w))$, is evaluated,

\begin{equation}
m_{\min }(\beta ,\phi ,\theta )=\frac{\cos ^{-1}(\sqrt{\frac{b-2a}{2b}})}{w}%
\text{,}
\end{equation}%
where

\begin{eqnarray*}
a &=&\sin (2x)\cos (2\beta )+\cos (2x)\cos (\frac{\phi }{2})\sin (2\beta ),
\\
b &=&(2+\sin ^{2}(2x)+(3\sin ^{2}(2x)-2)\cos (4\beta )-2\sin ^{2}(2x)\cos
(\phi )\sin ^{2}(2\beta )) \\
&&+2\sin (4x)\cos (\frac{\phi }{2})\sin (4\beta )\text{.}
\end{eqnarray*}

For a sure-success search problem, the phase condition, $\phi =\theta $,
provided iterations, $m_{\min }=(\pi /2-\sin ^{-1}(\sin (\phi /2)\sin (\beta
))/w$, is required. However, when effects of imperfect phase inversions are
considered, the search is not certain, then the new condition to phase
error, said $\delta =\phi -\theta $, and the size of database would be
rederived in order to accomplish the search with a reduced maximum
probability. Now, we suppose the database is large, i.e., if $\sin (\beta
)<<1$, and a phase error $\delta $ is small, where $\left| \delta \right|
<<1 $, one will have the following approximation, viz.,

\begin{eqnarray*}
\cos (w) &=&\cos (\frac{\delta }{2})-2\sin (\frac{\theta }{2}+\frac{\delta }{%
2})\sin (\frac{\theta }{2})\sin ^{2}(\beta ) \\
&\approx &1-(\frac{\delta ^{2}}{8}+2\beta ^{2}\sin ^{2}(\frac{\theta }{2}))%
\text{,} \\
\sin (w) &=&(1-\cos ^{2}(w))^{1/2} \\
&\approx &\frac{(\delta ^{2}+16\beta ^{2}\sin ^{2}(\frac{\theta }{2}))^{1/2}%
}{2}\text{,} \\
\sin (2x) &=&\frac{4\beta \sin (\frac{\theta }{2})}{(\delta ^{2}+16\beta
^{2}\sin ^{2}(\frac{\theta }{2}))^{1/2}}\text{.}
\end{eqnarray*}%
The probability $P$ , equation (9), then has the approximation

\begin{eqnarray}
P &\approx &1-\cos ^{2}(mw)\cos ^{2}(\beta )-\sin ^{2}(mw)\cos ^{2}(2x) \\
&=&\sin ^{2}(mw)\sin ^{2}(2x)\text{,}  \nonumber
\end{eqnarray}%
with a maximum value, by letting $\sin ^{2}(mw)=1$,

\begin{equation}
P_{\max }\approx \sin ^{2}(2x)=\frac{16\beta ^{2}\sin ^{2}(\frac{\theta }{2})%
}{\delta ^{2}+16\beta ^{2}\sin ^{2}(\frac{\theta }{2})}
\end{equation}%
The function (12) for two designations, $\delta =0.01$ and $\delta =0.001$,
are depicted in Fig. 1 and Fig. 2 respectively.

Observing Fig. 1 and Fig. 2, one realizes the function (12) depicted by
solid line coincides with the exact value, obtained by Eq. (9) and Eq. (10),
shown by cross marks. On the contrary, the result of Long et al.,

\begin{equation}
P_{\max }\approx \frac{4\beta ^{2}\sin ^{2}(\frac{\theta }{2})}{\delta
^{2}+4\beta ^{2}\sin ^{2}(\frac{\theta }{2})}\text{,}
\end{equation}%
is an underestimation depicted by dash lines.

To summarize, under the inevitable gate imperfections, we have derive the
generalized and improved criterion (12), by the exact formulation of Grover
kernel after $m$ numbers of iterations and the approximation to small values
of involved parameters, for tolerated error and its corresponding size of
quantum database. Moreover, the minimum number of iterations for obtaining
the maximum probability, $m_{\min }(\beta ,\phi ,\theta )$, is also
presented. By observing Fig.1 and Fig.2, one can realize the improved
criterion (12) is near an exact one. Besides, utilizing condition (12), one
can realize that the value of tolerated error decreases with the growth of
database, in other words, it is important to have a good control over
tolerated error if we have a large quantum database. Therefore, quantum
search machines should avoided gate imperfections as much as possible. If we
cannot get rid of these errors, we must limit the size of a quantum database
accurately. The result of this study presents a more accurate
characterization of the relation between systematic errors and the size of a
quantum database. A nearly exact criterion (12) can be utilized in order to
achieve the practical equilibrium between the actual gate imperfection and
the size of the quantum database.

\bigskip

\begin{itemize}
\item {\LARGE Figure Caption:}

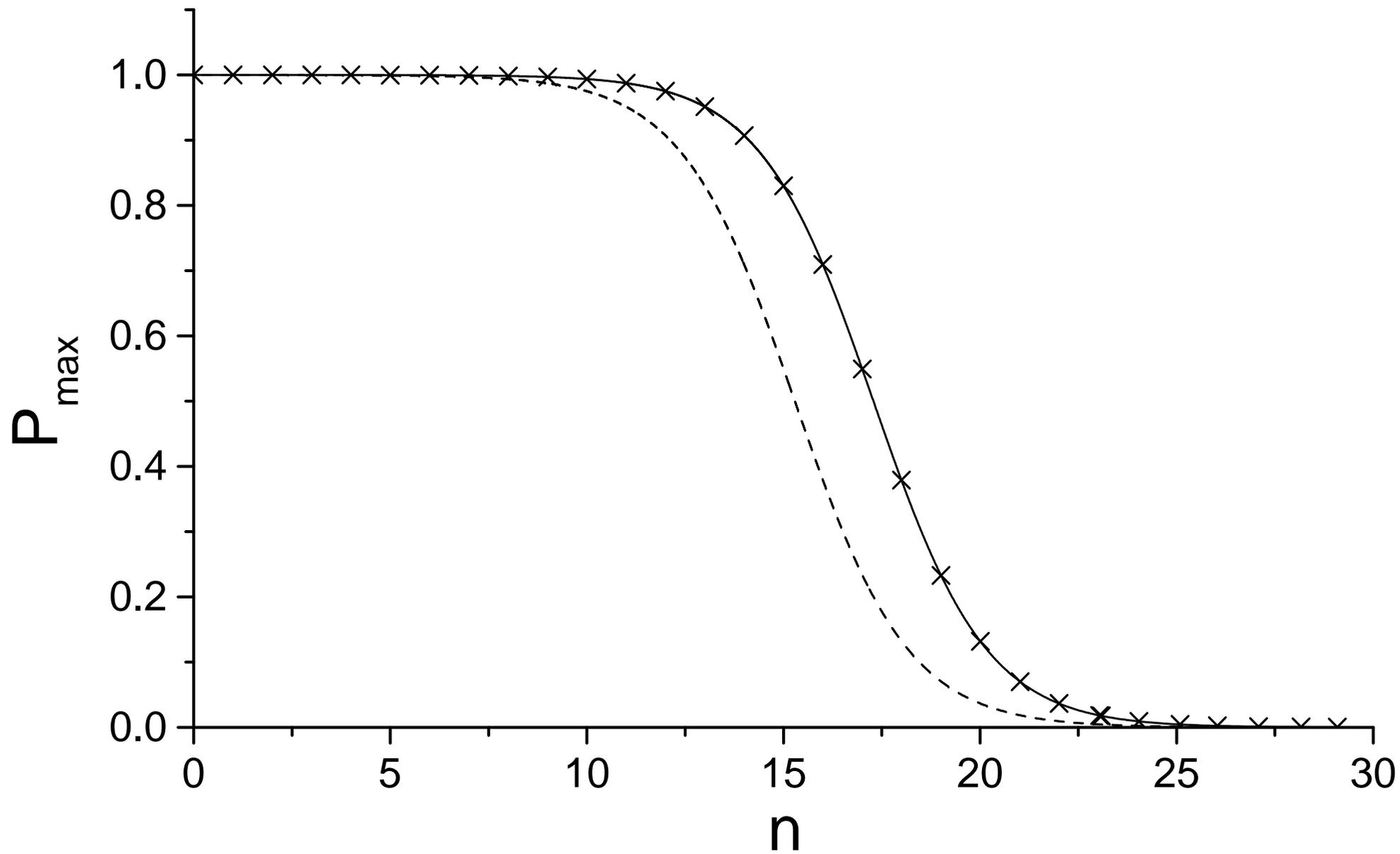
FIG. 1. Variations of exact value of $P_{\max }$ $(n)$(cross marks), $%
16\beta ^{2}\sin ^{2}(\frac{\theta }{2})/(\delta ^{2}+16\beta ^{2}\sin ^{2}(%
\frac{\theta }{2}))$ (solid), and $4\beta ^{2}/(\delta ^{2}+4\beta ^{2})$
(dash) for $\theta =\pi $, $\delta =0.01$ where $\beta =\sin ^{-1}(2^{-n/2})$%
.

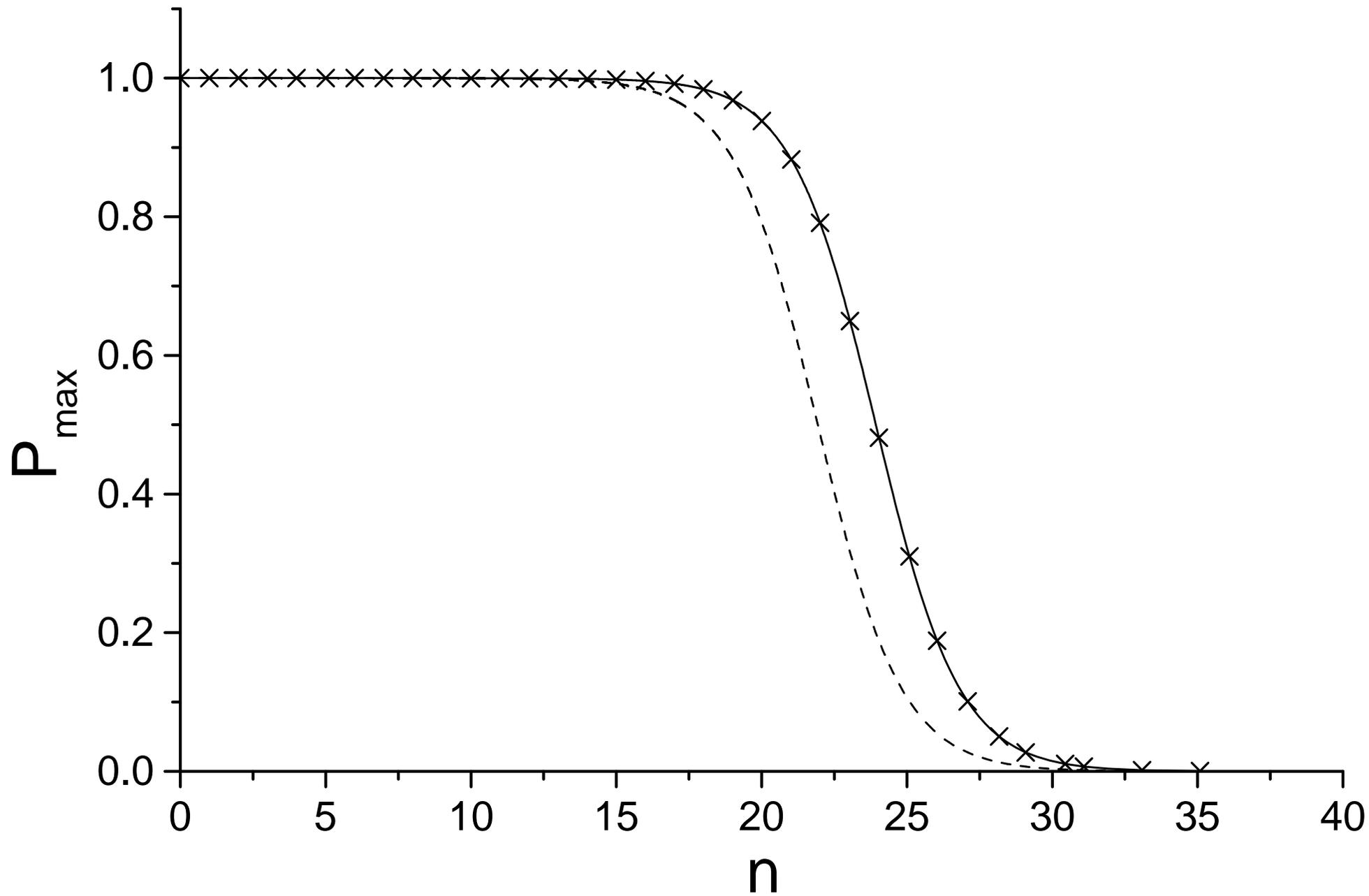
FIG. 2. Variations of exact value of $P_{\max }$ $(n)$(cross marks), $%
16\beta ^{2}\sin ^{2}(\frac{\theta }{2})/(\delta ^{2}+16\beta ^{2}\sin ^{2}(%
\frac{\theta }{2}))$ (solid), and $4\beta ^{2}/(\delta ^{2}+4\beta ^{2})$
(dash) for $\theta =\pi $, $\delta =0.001$ where $\beta =\sin
^{-1}(2^{-n/2}) $.
\end{itemize}

\end{document}